\documentclass[useAMS,usenatbib]{mn2e}
\usepackage{times}
\usepackage{epsfig}
\usepackage{amssymb}
\usepackage{amsmath} 
\usepackage{aas_macros}
\title[PDFs of line fluxes and ratios]{The probability distribution functions of emission line flux measurements and their ratios}
\author[R. Wesson et al.]
{R. Wesson$^1$, D.J. Stock$^2$, P. Scicluna$^3$\\
$^1$European Southern Observatory, Alonso de C\'ordova 3107, Casilla 19001, Santiago, Chile\\
$^2$Department of Physics and Astronomy, University of Western Ontario, London, ON N6K 3K7, Canada\\
$^3$Kiel University, Institute of Theoretical Physics and Astrophysics, Leibnizstr. 15, 24118 Kiel, Germany\\
}
\date{Received:}

\begin{document}
\maketitle

\begin{abstract}

Many physical parameters in astrophysics are derived using the ratios of two observed quantities.  If the relative uncertainties on measurements are small enough, uncertainties can be propagated analytically using simplifying assumptions, but for large normally distributed uncertainties, the probability distribution of the ratio become skewed, with a modal value offset from that expected in Gaussian uncertainty propagation.  Furthermore, the most likely value of a ratio A/B is not equal to the reciprocal of the most likely value of B/A.  The effect is most pronounced when the uncertainty on the denominator is larger than that on the numerator.

We show that this effect is seen in an analysis of 12,126 spectra from the Sloan Digital Sky Survey.  The intrinsically fixed ratio of the [O~{\sc iii}] lines at 4959 and 5007{\AA} is conventionally expressed as the ratio of the stronger line to the weaker line.  Thus, the uncertainty on the denominator is larger, and non-Gaussian probability distributions result.  By taking this effect into account, we derive an improved estimate of the intrinsic 5007/4959 ratio.  We obtain a value of 3.012$\pm$0.008, which is slightly but statistically significantly higher than the theoretical value of 2.98.

We further investigate the suggestion that fluxes measured from emission lines at low signal to noise are strongly biased upwards.  We were unable to detect this effect in the SDSS line flux measurements, and we could not reproduce the results of Rola and Pelat who first described this bias.  We suggest that the magnitude of this effect may depend strongly on the specific fitting algorithm used.

\end{abstract}

\begin{keywords}
atomic data -- methods: data analysis -- methods: statistical
\end{keywords}

\section{Introduction}

A great deal of information in the physical sciences is derived from the ratios of observed quantities.  The measurements of quantities are always associated with an uncertainty, and the uncertainty should of course be propagated into the resulting ratio.  When doing so, if the fractional uncertainty is small, one can use truncated Taylor expansions to derive approximate expressions for the uncertainties on derived quantities.  However, when dealing with real data, the fractional uncertainty is often not small, and biases may result from any invalid approximations made in propagating the uncertainties.

We outline in this paper some properties of the probability distributions of ratios which can significantly affect the interpretation of results when the signal to noise ratio of measurements is relatively low.  We first describe a number of mathematical axioms, and the biases which result from them, and then we show that these biases can be detected in observational data.  We derive an improved value of the intrinsic [O~{\sc iii}] 4959/5007 ratio by ensuring that the biases described are minimised.  We then consider whether line flux measurements at low signal to noise ratios are strongly biased upwards, as has previously been suggested.  Finally we discuss the circumstances in which these biases may lead to erroneous conclusions.

\section{The probability distribution of the ratio of Gaussians}

If two quantities X and Y have independent Gaussian probability density functions, both with mean zero and variance of unity, then the probability distribution of their ratio X/Y, f(X/Y), has a Lorentz distribution (\citealt{Marsaglia:1964:RNV}, \citealt{HINKLEY01121969}, \citealt{Marsaglia:2006:RNV}):

\begin{equation}
f(X/Y) \propto \frac{1}{1+x^2}
\end{equation}

The ratio of two non-zero quantities with Gaussian probability distributions has a probability density function which is Lorentz-like, but cannot be expressed in closed form (\citealt{HINKLEY01121969}).  It can be approximated by a Gaussian distribution when the fractional uncertainty is small, but becomes increasingly skewed as the fractional uncertainties increase.  The mean of this ratio distribution is mathematically undefined and the mode of the distribution is not equal to the ratio of the modes of the two quantities.  Also, for two quantities $X$ and $Y$, the mode of the probability distribution of $X/Y$ is not the same as the reciprocal of the mode of the distribution of $Y/X$.

\begin{figure}
\includegraphics[width=0.5\textwidth]{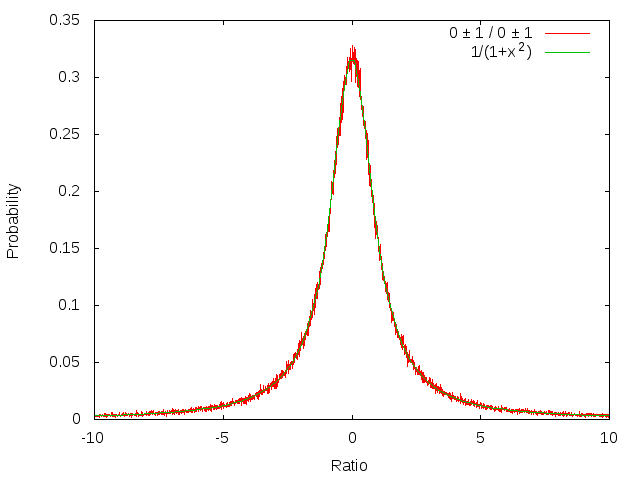}
\caption{The distribution of the ratios of two normally distributed variables is a Lorentz distribution.}
\label{lorentz}
\end{figure}

We have carried out Monte Carlo analyses to illustrate these axioms.  Firstly, we drew pairs of independent random numbers from a Gaussian distribution with mean of zero and variance of unity and took their ratio.  Figure~\ref{lorentz} shows the distribution of 1\,000\,000 such ratios, together with a Lorentz distribution having $\gamma$=1.0 and $x_0$=0.

\begin{figure}
\includegraphics[width=0.5\textwidth]{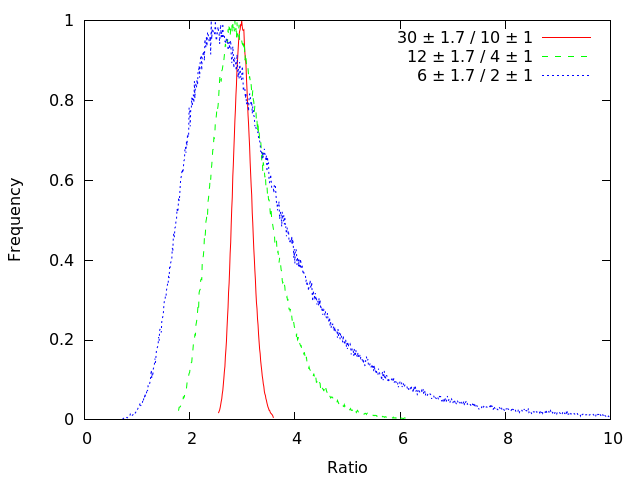}
\caption{Scaled probability density functions for the ratios of non-zero normally distributed variables.}
\label{generalratio}
\end{figure}

Figure~\ref{generalratio} shows the results of three further Monte Carlo simulations, this time for the ratios of variables with normal distributions and non-zero means.  We drew random numbers from Gaussian distributions with means of 30 and 10; 12 and 4; and 9 and 3.  The standard deviations of the distributions were 1.7 and 1.0 in each case to simulate the common case of Poissonian noise.  Thus, the ratio of the quantities is 3.0 but the fractional uncertainty varies.  The figure shows the probability distributions scaled such that the peak is at unity in each case; it can be seen that when the fractional uncertainty is small, the mode of the probability distribution of the line ratio is very close to 3.0, and the distribution is very close to Gaussian.  But as the fractional uncertainty increases, the skew of the distribution increases and the mode is offset to lower values.  For the case where the signal to noise ratios of the two values are 6 and 2, the mode of the resulting probability distribution is 2.3.

\begin{figure*}
\includegraphics[width=0.45\textwidth]{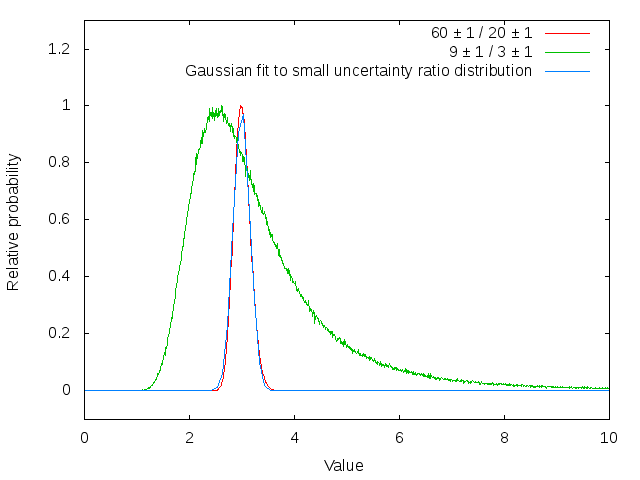}
\includegraphics[width=0.45\textwidth]{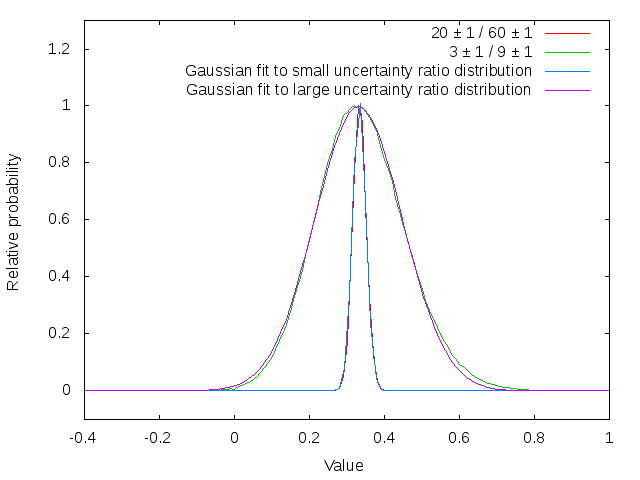}
\caption{Scaled probability density functions for ratios where the denominator has the larger uncertainty (l) and where the numerator has the larger uncertainty (r).}
\label{ratiosandinverses}
\end{figure*}

Figure~\ref{ratiosandinverses} shows two sets of probability distributions, both derived from the same set of 1,000,000 pairs of randomly chosen numbers with Gaussian uncertainties.  As before, the distributions are scaled such that the peak is at unity for ease of comparison.  The first shows the probability distribution of the ratio of the larger quantity to the smaller quantity, for large and for small uncertainties.  When the uncertainties are small (5 per cent on the smaller quantity in this example), the ratio distribution is approximately Gaussian, though a divergence from Gaussian can already be seen when a Gaussian function derived by non-linear least squares fitting to the observed distribution is overplotted.  When the uncertainties are large, the ratio distribution is highly non-Gaussian.  The second panel of the figure shows the probability distributions of the ratio of the smaller quantity to the larger quantity, and in this case it is apparent that even when the uncertainties are large, the ratio distribution is still fairly well approximated by a Gaussian distribution.

We thus summarise by making the following general points about the probability distributions of ratios:

\begin{itemize}
\item The probability distribution of the ratio of two Gaussian variables is not Gaussian, but can in certain circumstances be approximated as such;
\item when the uncertainty on the denominator is smaller than that on the numerator, a Gaussian approximation is reasonable even when the uncertainty is relatively large;
\item when the uncertainty in the denominator is larger than that on the numerator, a Gaussian approximation is not reasonable even when the uncertainty is relatively small;
\item when the ratio distribution is non-Gaussian, the mode of its probability distribution is not equal to the ratio of the modes of the input Gaussian distributions.
\end{itemize}

These points result from the probability of numbers close to zero becoming more probable as the uncertainties increase.  When the denominator has large uncertainties and thus a significant probability of being close to or less than zero, the probability distribution of the ratio becomes highly non-Gaussian with a heavy tail and a significant probability of very large values.  As the uncertainty of the denominator tends to zero, the probability distribution of the ratio tends towards a simple scaling of that of the numerator.

For the skewed distributions resulting from large uncertainties on the denominator, the skew is always to the right.  Taking the mode of a sample of ratios in which the uncertainty of the denominator is large will lead to an underestimate of the true value of the ratio.  Considering the mean, at intermediate signal to noise ratios it will give an overestimate of the true value of the ratio due to the right hand skew of the distribution, but as the signal to noise of the denominator tends towards zero, the distribution tends towards a Cauchy distribution, in which the mean is undefined and the probability distribution of a sample mean is the same as that of an individual sample.  The median is almost unbiased except when the signal to noise ratio of the denominator is very low and the distribution becomes significantly bimodal.

We now consider the extent to which the effects described are observationally relevant.

\section{Evidence of biases in observational datasets}

One way to detect observational biases is to examine emission lines which have a fixed intrinsic ratio, are closely spaced in wavelength to avoid undue influence of systematic effects such as uncertainties in the flux calibration and correction for interstellar reddening, and which are detected in a large number of astronomical spectra.  Several such sets of lines exist: the nebular lines of [O~{\sc iii}] at 4959 and 5007 {\AA}, [N~{\sc ii}] at 6548 and 6584 {\AA}, and [O~{\sc i}] at 6300 and 6363 {\AA} all arise from a common upper level, and thus their ratio is intrinsically fixed.  These line pairs are widely observed, with theoretical ratios of 2.98 ([O~{\sc iii}]), 3.01 ([N~{\sc ii}]) and 2.97 ([O~{\sc i}]) (\citealt{2000MNRAS.312..813S}).  The [O~{\sc iii}] and [N~{\sc ii}] line pairs are among the brightest lines emitted at optical wavelengths by the gas ionised by hot stars and so are widely detected; the [O~{\sc i}] lines are typically much weaker and are thus measured with much lower signal to noise ratios.

Our work on biases affecting line ratios was originally intended to identify observational evidence of the effect described by \citet{1994A&A...287..676R}, in which line intensities are much more likely to be overestimated than underestimated at low signal to noise, and the probability distribution describing the measurement becomes better described by a log-normal distribution than a normal distribution.  In our code {\sc neat}, described in \citet{2012MNRAS.422.3516W}, we use a Monte Carlo technique to propagate uncertainties, which allows non-Gaussian uncertainties to be straightforwardly propagated, and we use equations derived in \citet{1994A&A...287..676R} to determine the appropriate log-normal distribution to adopt when the signal to noise ratio of a given line measurement is low.  We found that assuming log-normal probability distributions resulted in a smaller estimated uncertainty on quantities derived from weak lines.

However, the existence of this bias affecting weak lines has been questioned; \citet{2013A&A...552A..12S}, for example, discussed the effect of non-Gaussian probability distributions, and said that ``{\it the strong biases} [\citet{2012MNRAS.422.3516W}] {\it claim for line intensities with signal-to-noise ratios $\leq$4 are not supported by observations of line ratios such as [O~{\sc iii}] $\lambda$4959/5007 or [N~{\sc ii}] $\lambda$6548/6484 which, on average, are consistent with the values predicted by atomic physics (see, e.g., Fig.  10 of \citealt{2005A&A...441..981B}})". We therefore sought to investigate whether biases at low signal to noise ratios do indeed exist in observational datasets.

We first looked at the data in \citet{2005A&A...441..981B}, which presents emission line fluxes for 70 extragalactic H~{\sc ii} regions.  Of these 70, the [O~{\sc iii}] and [N~{\sc ii}] lines are detected in 68 objects.  Averaged over the sample, the ratios are 2.91$\pm$0.57 for [O~{\sc iii}] (2.85$\pm$0.29 if one point where the ratio is 7.0 is excluded), and 2.97$\pm$0.14 for [N~{\sc ii}].  These averages are thus consistent with the values predicted by atomic physics, as stated by \citet{2013A&A...552A..12S}.  However, when we consider the behaviour of the estimated line ratio with signal to noise ratio, we find that the ratios at low signal to noise systematically differ from the predicted value.

\begin{figure}
\includegraphics[width=0.5\textwidth]{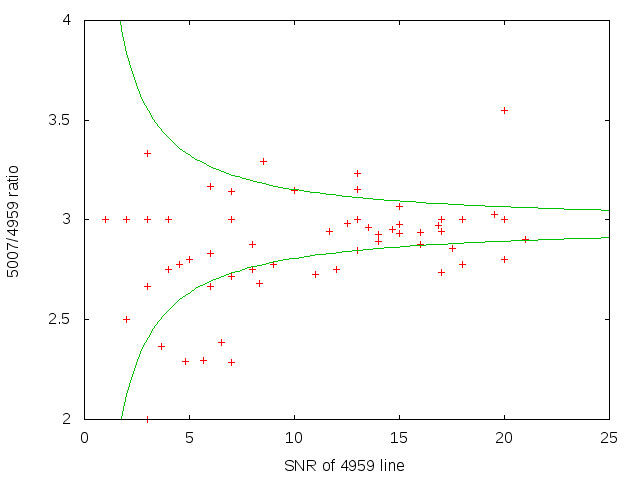}
\caption{Ratio against signal to noise for the dataset of \citet{2013A&A...552A..12S}}
\label{bresolin-snr-figure}
\end{figure}

In Figure~\ref{bresolin-snr-figure}, we plot the observed ratio of these lines against the SNR for the weaker line, together with an estimate of the 1$\sigma$ uncertainty bounds using standard analytical uncertainty propagation equations.  In the case of the [O~{\sc iii}] lines, the reported signal to noise ratio of the weaker line is lower than 6.0 in 26 objects, while for the [N~{\sc ii}] lines, the lowest SNR in the sample is 7.0.  The [O~{\sc iii}] lines show a clear tendency for the line ratio to be lower than its theoretical value, a tendency which is not seen in the case of the [N~{\sc ii}] lines.  Table~\ref{bresolin-snr-table} shows the mean line ratio from the values in various ranges of SNR, for both species.

\begin{table}
\begin{tabular}{lll}
\hline
SNR range & [O~{\sc iii}] ratio & [N~{\sc ii}] ratio \\
\hline
0-3 & 2.71 & -- \\
3-6 & 2.74 & --\\
6-10 & 2.85 & 2.96 \\
10-20 & 2.96 & 2.96 \\
$>$20 & 3.06 & -- \\
\hline
\end{tabular}
\caption{The mean value of line ratios in SNR bins in the data of \citet{2013A&A...552A..12S}}
\label{bresolin-snr-table}
\end{table}

If the variance in the \citet{2005A&A...441..981B} dataset could be well described by Gaussian statistics, approximately 16\% of values should lie outside the upper 1$\sigma$ limit, and 16\% below the lower 1$\sigma$ limit.  For 68 data points, this would imply 11 either side.  In fact, 16 lie below the lower limit and only 4 above the upper limit.  We estimated the probability of this occurring by chance if the data were described by a Gaussian probability distribution by carrying out a Monte Carlo simulation, and found that for 68 samples drawn from a Gaussian distribution, finding 16 or more samples more than 1$\sigma$ below the mean and 4 or less samples more than 1$\sigma$ above the mean occurs about 0.2\% of the time.

We next sought evidence of the effect in a much larger dataset.  We used emission line flux measurements from the Sloan Digital Sky Survey 9th Data Release (\citealt{2012ApJS..203...21A}), which contains emission line fluxes for 2,674,203 objects.  Of the three line pairs mentioned earlier, only the [O~{\sc iii}] 5007/4959 ratio is available; of the [O~{\sc i}] lines, only 6300{\AA} fluxes are given, and for the [N~{\sc ii}] lines, the line fluxes are not measured independently, the ratio instead being fixed to its expected theoretical value of 3.01.  We thus consider the [O~{\sc iii}] 5007/4959 ratio only.

Interstellar reddening will increase this ratio above its theoretical value, but the small separation in wavelength of these two lines means that the effect is slight.  Using the extinction curve of \citet{1983MNRAS.203..301H}, the difference in the value of f($\lambda$) is 0.012, so that for a c(H$\beta$) of 1.0, the ratio would be increased by 3 per cent.  Meanwhile, the statistical effect described above will reduce the modal value of the observed ratio.  To minimise the confounding influence of reddening, we extracted from the DR9 table of emission line fluxes those objects for which c(H$\beta$) calculated from the observed H$\alpha$/H$\beta$ ratio was less than 0.05.  In this case, the ratio would be within 0.1\% of its predicted value if affected only by reddening.

\begin{figure*}
\includegraphics[width=\textwidth]{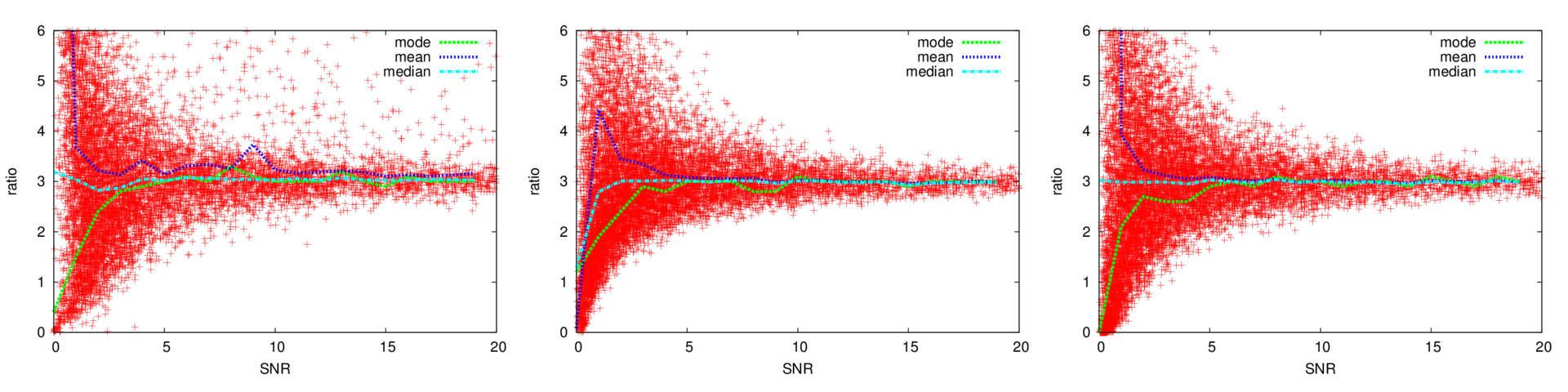}
\caption{The ratio of the [O~{\sc iii}] lines at 4959 and 5007 {\AA} as a function of signal to noise ratio.  The left panel shows the relation derived from 12,126 SDSS spectra.  The centre panel shows the relation in a Monte Carlo simulation using normally distributed variables described in the text, while the right panel shows the results for log-normally distributed variables.  In all panels, the modal value of the ratio in uncertainty bins of unit width is overplotted.}
\label{sdss-histogram}
\end{figure*}

Restricting our data set to those objects where c(H$\beta) < $0.05, and where the value of c($H\beta$) has an uncertainty of less than ten per cent, we are left with 12,126 measurements of the [O~{\sc iii}] line ratio.  Figure~\ref{sdss-histogram} shows a plot of the line ratio against the reciprocal of the uncertainty of the ratio, calculated from the quoted uncertainties on each line flux measurement.

Figure~\ref{sdss-histogram} shows on the left hand panel the relation between observed ratios and the reciprocal of the estimated uncertainty for SDSS line flux ratios.  The centre panel shows the relation between the calculated ratio and the reciprocal of its uncertainty for a Monte Carlo simulation in which 12,126 pairs of numbers were created as follows: first, a value $S$ representing the signal to noise ratio was taken from a log-normal distribution with $\mu$=1.0 and $\sigma$=1.3; these values were chosen to approximate the distribution of observed signal to noise ratios in the SDSS data set.  Then, two random numbers representing line fluxes were chosen, one from a normal distribution with $\mu$=1.0 and $\sigma$=($\sqrt{3}/2S$), and the second from a normal distribution with $\mu$=3.0 and $\sigma$=$1/2S$, such that standard analytical uncertainty propagation results in an estimated uncertainty on the line ratio of $3.0/S$.

There is a strong resemblance between the two plots, with both clearly showing modal values of the ratio being biased at low signal to noise.  However, some differences are clear, with a far larger number of outlying points in the observed data, and a greater excess of values below the theoretical ratio at low signal to noise.  A possible cause of at least part of the difference is that the line fluxes of forbidden lines at low signal to noise ratios cannot be represented by a Gaussian probability distribution, since physically, these lines cannot exist in absorption.  At low SNR, a Gaussian probability distribution would imply an unphysical non-zero probability of negative line flux.  We therefore consider how non-Gaussian probability distributions would affect these results.

\subsection{The probability distribution of the ratio of truncated Gaussian distributions}

We first consider the probability density function of the ratio of Gaussian distributions truncated at zero.  We find that the probability of the ratio being negative is then, of course, zero, but that the behaviour of the mode of the probability distribution of the ratio is not significantly affected by the truncation, as it arises from the effect of the reciprocals of very small numbers, which are still present in the probability distributions.

\subsection{The probability distribution of the ratio of log-normal distributions}

As mentioned, the probability distribution of the 4959 and 5007{\AA} line fluxes at low SNR cannot be Gaussian.  In addition, the number of outliers at large multiples of the standard deviation in the observed line ratio must indicate either than the probability distributions are not Gaussian, or that the uncertainties are underestimated, or both.  \citet{1994A&A...287..676R} identified an effect which would give rise to non-Gaussian probability distributions for measured line fluxes and skew line ratios at low signal to noise.  They found that for narrow lines in noisy spectra, flux measurements become much more likely to be overestimated than underestimated, and that their probability distribution can be best described by a log-normal distribution, becoming increasingly skewed as the signal to noise ratio decreases.  Log normal distributions maintain the criterion that negative fluxes should have a probability of zero, and also have the useful property that the probability distribution function of the ratio of two log-normally distributed variables is itself log-normal.  

To investigate the ratios of log-normal distributions, we carried another Monte Carlo simulation similar to that described above for normal ratios, but in which the fluxes were drawn from log-normal distributions, with $\mu$=log(1.0) and $\sigma$=($\sqrt{3}/2S$), and $\mu$=3.0 and $\sigma$=$1/2S$.  The results are plotted in the right hand panel of Figure~\ref{sdss-histogram}.  We find that for the ratios of log-normally distributed variables, the mode and mean are biased in a similar way to that of normally distributed variables, while the median remains an unbiased estimator regardless of signal to noise.  Visually, the scatter plot of ratio against SNR appears to be more similar to the observed distribution, having a ``bulge" of low ratios at low signal to noise.

These log-normal distributions only imply asymmetric uncertainties, and not a systematic overestimate of weak line fluxes as envisaged by \citet{1994A&A...287..676R}.  We thus note that the observed distribution of ratios with signal to noise in the SDSS data can be broadly reproduced whether the synthesized line fluxes are normally or log-normally distributed, without the need to invoke any systematic overestimate of line fluxes at low signal to noise ratios.

\subsection{Are line flux measurement uncertainties Gaussian at low signal to noise ratios?}

Having found that it is not necessary to invoke an upward bias in line flux meeasurements to account for the skewedness of the probability distributions of ratios such as observed in the SDSS data, we then sought to replicate the results of \citet{1994A&A...287..676R}.  We carried out a Monte Carlo simulation in which we created several thousand synthetic spectra, consisting of a Gaussian profile with $\sigma$=1.0 and peak=1.0, superimposed on a continuum created using random numbers symmetrically distributed about zero such that the expected value of the sum of the continuum points was zero.  We then used a non-linear least squares fitting routine to attempt to determine the parameters of the line by fitting a Gaussian profile to the synthetic spectrum.  The initial guesses for the Gaussian parameters were set to be similar but not equal to their true values.

We investigated a number of different approaches.  Firstly, we followed the methodology described in \citet{1994A&A...287..676R}, in which four ``spectra" are created, one of which contains a line and the other three are pure noise.  We then first attempt to detect which sample contains the line, and then try to fit a Gaussian function to that sample.  We also investigated the case in which a line was always present, such that only false negatives were possible, and not false positives.  In both of these cases, we also investigated continua created from normally distributed and uniformly distributed random numbers.  In the case of creating four spectra, our code calculated the variance of each spectrum and assumed that the spectrum with the largest variance was the one containing the line.  In all cases, fits were rejected either when the non-linear least squares routine failed to converge, or when it reported negative peaks or sigmas.

In the ``four samples" case, we found that with decreasing SNR there was in fact a downward trend in the median estimated flux of the line; the false positives that the routine successfully fitted a Gaussian profile to were typically noise features one or two pixels wide, while for $\sigma$=1.0, the full width at half maximum of a Gaussian profile is 2.35 pixels.  Thus, the reported $\sigma$ of spurious features was generally less than 1.0 and the reported flux lower than the actual value.  In real spectral fitting one would be able to identify such false positives, as long as the intrinsic line profile was not significantly undersampled by the detector.  The noise in our synthetic continua was uncorrelated, and noise with non-negligible auto-correlation could give rise to a larger bias as noise peaks several pixels wide would then be more probable.

In the ``one sample" case, we found that with decreasing SNR, there was no trend in the estimated line centre, a downward trend in the estimated line peak, and an upward trend in the estimated line width.  The trends in the width and peak almost cancelled, such that there was only a very small upward trend in the estimated flux.  In both cases, we found that the probability distribution of the measured flux could be approximated by a truncated Gaussian distribution at low signal to noise ratios.

\begin{figure*}
\includegraphics[width=0.45\textwidth]{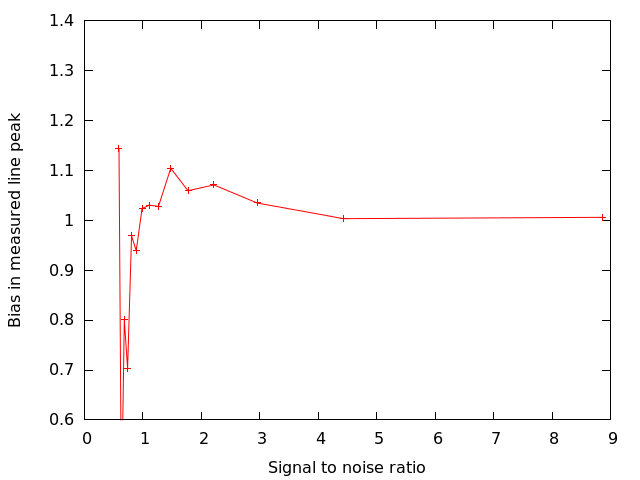}
\includegraphics[width=0.45\textwidth]{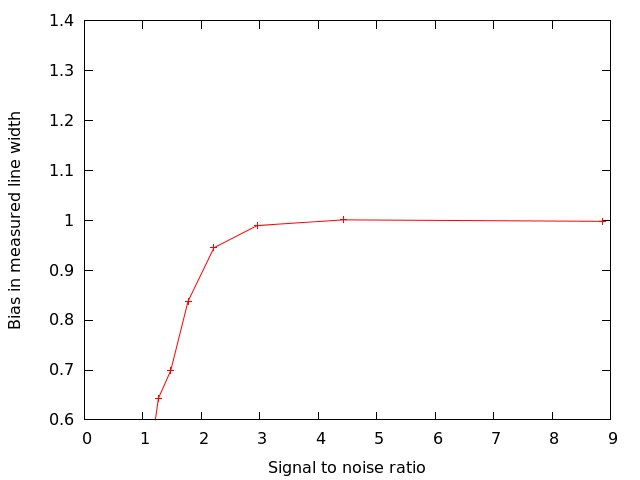}
\includegraphics[width=0.45\textwidth]{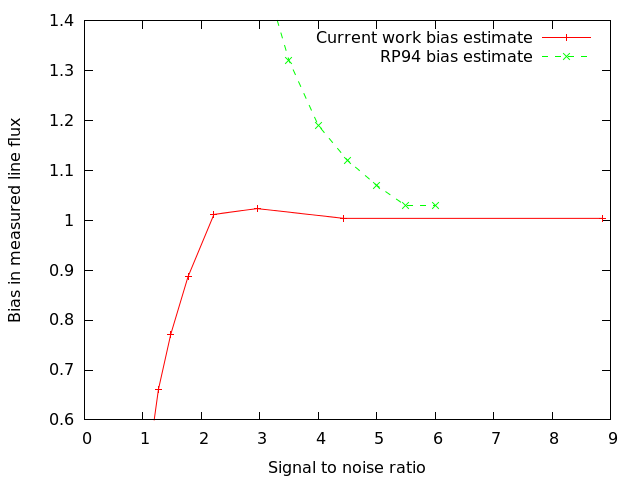}
\includegraphics[width=0.45\textwidth]{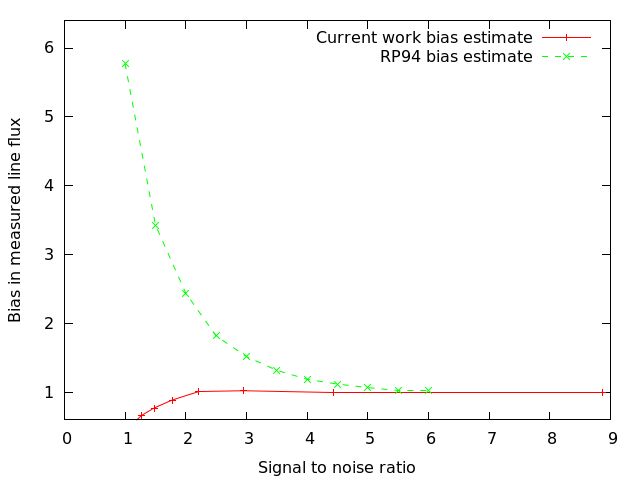}
\caption{Gaussian parameters estimated by non-linear least squares fitting for synthetic spectra containing a Gaussian line profile with width and peak of 1.0, and thus a flux of $\sqrt{2\pi}$, superimposed on a white noise continuum.  This figure shows the results for a fitting procedure which generated both false positives and false negatives.  At low signal to noise ratios, a downward bias in line flux measurements is present.}
\label{RPreplication}
\end{figure*}

\begin{figure*}
\includegraphics[width=0.45\textwidth]{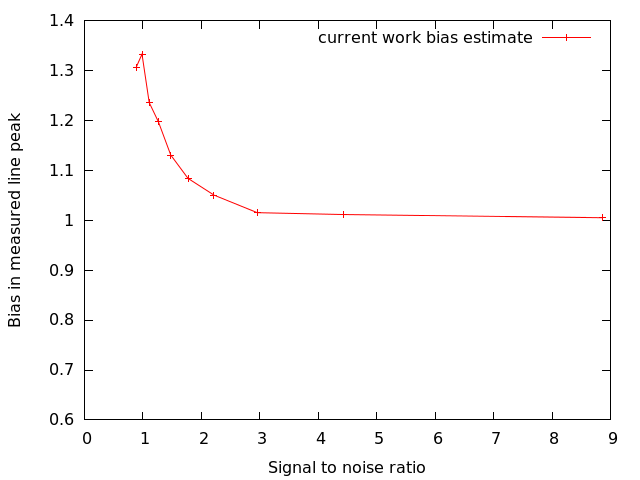}
\includegraphics[width=0.45\textwidth]{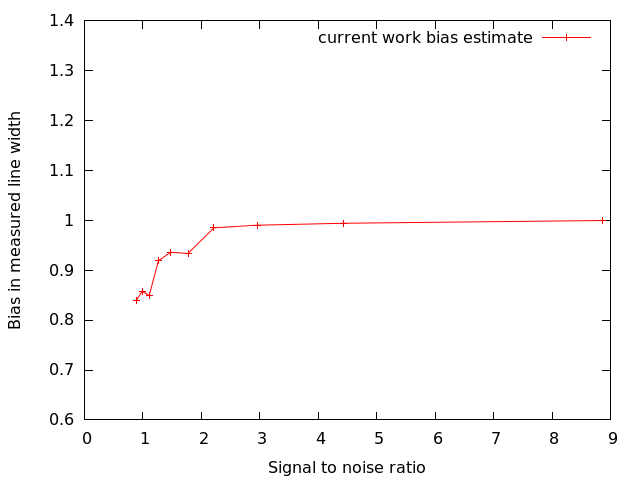}
\includegraphics[width=0.45\textwidth]{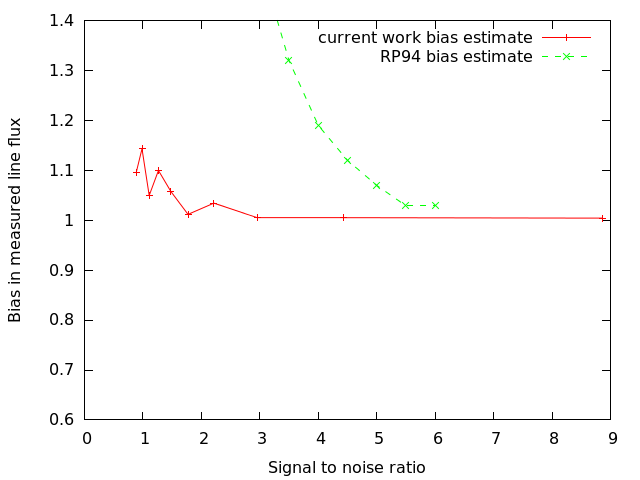}
\includegraphics[width=0.45\textwidth]{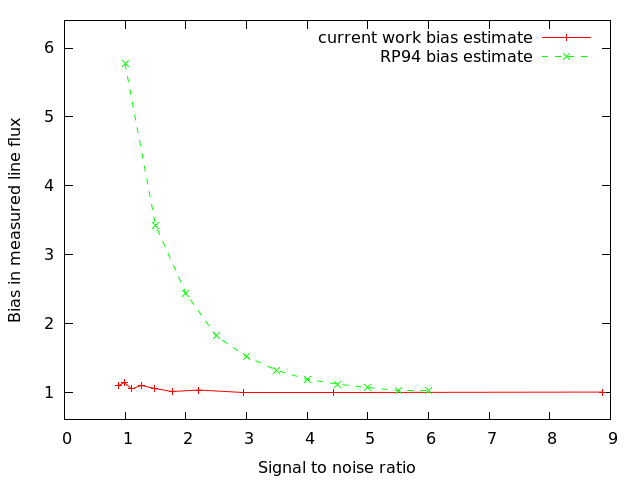}
\caption{Same as Figure~\ref{RPreplication} but for a fitting procedure in which a line was always present and thus only false negatives were possible.  In this case, an upward bias is seen in flux measurements but to a far lesser degree than that found by \citet{1994A&A...287..676R}.}
\label{RPreplication2}
\end{figure*}

We suspect therefore that the considerable effect found by \citet{1994A&A...287..676R} must be strongly dependent on the approach used to determine whether a line is present or not, and also on the optimisation algorithm used to fit the Gaussian.  The widely used Marquardt-Levenburg algorithm, which we used in this experiment, seems to be robust against the overestimation of line fluxes at low signal to noise ratios.  Our results which show the bias to be small or even opposite to that found by \citet{1994A&A...287..676R} are displayed in Figures~\ref{RPreplication} and \ref{RPreplication2}.

We conclude, therefore, that while the SDSS line flux measurements and their quoted uncertainties provide clear evidence of a skewed probability distribution, there is no clear evidence of any inherent upward bias in the measurement of weak lines.  The scatter seen in Figure~\ref{sdss-histogram} is likely to be due to misestimation of uncertainties rather than effects of non-Gaussian line fluxes.

\section{Discussion}

We have described the probability distributions of ratios when uncertainties are large, which can be strongly non-Gaussian when the uncertainty on the denominator is larger than that on the numerator, but are markedly less so when the opposite is true.  Thus, Gaussian statistics can be assumed by ensuring that, when taking ratios of quantities, the uncertainty of the denominator is small.

\subsection{The intrinsic ratio of the [O~{\sc iii}] nebular lines}

The fact that the [O~{\sc iii}] nebular lines have a fixed intrinsic ratio allowed us to investigate the probability distributions of the ratios of noisy measurements.  We can now reinvestigate the value of that fixed intrinsic ratio.  Its value has been estimated on a number of occasions and surprising variation has been found, which has sometimes been taken to suggest a sizable discrepancy between the observed value and the best theoretical value of 2.98 (\citealt{2000MNRAS.312..813S}).  Previous reported values include 3.02$\pm$0.03 (\citealt{1985Msngr..39...15R}); 3.17$\pm$0.04 (\citealt{1987A&A...186...84I}); 3.00$\pm$0.08 (\citealt{1996A&AS..116...95L}); and 2.993$\pm$0.014 (\citealt{2007MNRAS.374.1181D}).  These values were derived from observations of a sample of H~{\sc ii} regions, spatially resolved spectra of a starburst galaxy, a sample of planetary nebulae, and a sample of active galactic nuclei respectively.  The authors who derived values from samples of objects all took the mean of their values as the final observed value; in the presence of non-Gaussian probability distributions this can be a biased estimator of the true value.

From the 12,126 SDSS spectra in which we found the reddening to be well determined and low, such that the effect of reddening on the observed ratio would amount to less than 0.1 per cent, we calculated both the 5007/4959 and 4959/5007 ratios.  As discussed above, the 5007/4959 ratio demonstrates clear non-Gaussian effects, and the mode, mean and median of the values diverge significantly at low signal to noise.  From 3316 values where the uncertainty on the ratio is less than 10 per cent, we find a median ratio of 3.012.  The mean and standard deviation of the 3316 values is 3.10$\pm$0.50, while the mode is 2.97, the divergence of these three parameters indicating that the probability distribution is not Gaussian.  For the 4959/5007 ratio, though, the mean, mode and median are all very similar even at low signal to noise ratios, indicating that the probability distribution of this ratio can be well approximated as Gaussian.  For the ratios where the estimated uncertainty is less than 10 per cent, we derive a median value of 0.332, corresponding to a 5007/4959 ratio of 3.012.  We estimated the uncertainty of these medians using a bootstrapping technique in which we randomly selected 10 per cent of the data points and determined the median of the subsample, repeating the process 100 times.  The resulting mean and standard deviations of the medians were 3.012$\pm$0.008 and 0.332$\pm$0.001.  We thus conclude that the observed ratio is slightly higher than the theoretical ratio, and that though the discrepancy is small at just one per cent, it is statistically significant, amounting to a 4$\sigma$ deviation, assuming that no other observational biases are present and neglecting any uncertainty on the theoretical value.


\section{Conclusions}

The uncertainty of the ratio of normally distributed variables is not itself normally distributed.  We have shown that it can be well approximated as such only when the uncertainty on the denominator is smaller than that of the numerator.  In the contrary case, the distribution becomes skewed, with the most likely value of the ratio being lower than the true ratio.  We have shown that this effect is detected in SDSS spectra, using the intrinsically fixed ratio of the [O~{\sc iii}] 4959 and 5007 lines.  The 5007/4959 ratio is biased by the generally larger uncertainty on the 4959 line, but the 4959/5007 ratio is not significantly biased.  We have used that fact to determine an estimate of the intrinsic line ratio with a statistical uncertainty of 0.25 per cent, sufficient to determine that the theoretical value differs by about 1 per cent from the observed value.

Many ratios are conventionally expressed such that the uncertainty on the denominator is typically larger, and these ratios will have non-Gaussian probability distributions even at quite high signal to noise ratios.  Using the mean or mode of a sample of ratios as an estimator of the true value, such as has been done in the past with the 4959/5007 ratio, will lead to incorrect results.  Ensuring that the numerator of a ratio has the larger uncertainty results in probability distributions which can be much better approximated as Gaussian.

\section{Acknowledgements}

We thank the anonymous referee for their helpful report.  This work was co-funded under the Marie Curie Actions of the European Commission (FP7-COFUND).  DJS acknowledges support from an NSERC Discovery Grant and an NSERC Discovery Accelerator Grant.  PS is partially supported under DFG programme no. WO 857/10-1.  RW thanks Mirjam Boere for helpful discussions.

\bibliographystyle{../mn2e}
\bibliography{statsbiases}

\end{document}